\documentclass[aps,prb,amsmath,amssymb,floatfix,twocolumn]{revtex4-1}
\usepackage{graphicx}
\usepackage{subfigure}
\usepackage{color}

\usepackage{hyperref}

\begin{document}

\title{Topological defects of N\'eel order and Kondo singlet formation \\
for Kondo-Heisenberg model on a honeycomb lattice}

\author{Pallab Goswami}
\affiliation{National High Magnetic Field Laboratory and Florida State University, Tallahassee, Florida 32310, USA}
\author{Qimiao Si}
\affiliation{Department of Physics and Astronomy, Rice University, Houston, TX 77005, USA}

\begin{abstract}
Heavy fermion systems represent a prototypical setting to study magnetic quantum phase transitions.
A particular focus has been on the physics of Kondo destruction, which captures quantum criticality
beyond the Landau framework of order-parameter fluctuations. In this context,
we study the spin one-half  Kondo-Heisenberg model  on a honeycomb lattice at half filling.
The problem is approached from the Kondo destroyed, antiferromagnetically ordered insulating phase.
We describe the local moments in terms of a coarse grained quantum non-linear sigma model,
and show that the skyrmion defects of the antiferromagnetic order parameter host
a number of competing order parameters. In addition to the spin Peierls, charge and current density wave
order parameters, we identify for the first time Kondo singlets as the competing orders of the antiferromagnetism.
We show that the antiferromagnetism and various competing singlet orders can be related to each other
via generalized chiral transformations of the underlying fermions.
We also show that the conduction electrons acquire a Berry phase through their coupling to the hedgehog configurations
of the N\'eel order, which cancels the Berry phase of the local moments. Our results demonstrate the competition
between the Kondo-singlet formation and spin-Peierls order when the antiferromagnetic order is suppressed,
thereby shedding new light on the global phase diagram of heavy fermion systems at zero temperature.

\end{abstract}

\maketitle

\section{Introduction}
\begin{figure}[htb]
\includegraphics[scale=0.4]{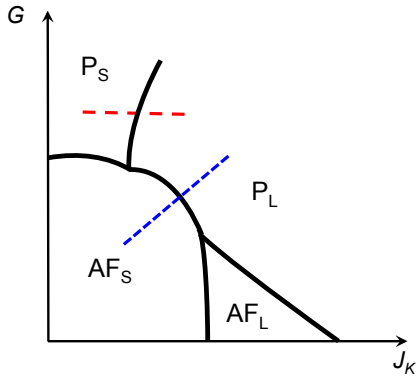}
\caption{(Color online) A proposed global phase diagram of the Kondo Heisenberg model in the presence
of the magnetic frustration $G$ and the Kondo coupling $J_K$
(Ref.~\cite{Si_PhysicaB2006}). Inside the paramagnetic ($P$)
and the antiferromagnetic ($AF$) phases respectively denoted by $P_S$ and $AF_S$, there is
no static Kondo screening; they represent Kondo destroyed phases.
The paramagnetic heavy Fermi liquid phase is represented by $P_L$. Both the AF order
and the static Kondo screening are present inside the phase $AF_L$,  which can be considered
as the spin density wave ordering of the heavy Fermi liquid quasiparticles.
In the present work we access the magnetically disordered phases starting from the $AF_S$ phase,
as illustrated by the blue dashed line. In this way, we study the competition between
$P_S$ and $P_L$; the red dashed line illustrates a possible path of phase transition between these phases.}
\label{Fig1}
\end{figure}

The competition between the local moment antiferromagnetism and the Kondo singlet formation
is a quintessential feature of the phase diagrams of many heavy fermion compounds \cite{SiSteglich,Lohneysen_rmp}.
This competition is responsible for the multitude of quantum critical points and emergent phases.
General considerations of such a competition have led to the theoretical proposal
for a global phase diagram (Fig.~\ref{Fig1})
\cite{Si_PhysicaB2006, YamamotoSi_JLTP2010,Coleman_JLTP2010},
which delineates the interplay between the the destruction of the static Kondo screening and the
onset of the local moment's anti-ferromagnetic (AF) order.
This global phase diagram has provided the understanding of the zero-temperature phase diagram in YbRh$_2$Si$_2$
in the multi-parameter space of  magnetic field, pressure, and chemical doping \cite{Friedemann09,Custers10,Tokiwa09},
and motivated the exploration of quantum phase transitions in heavy-fermion materials with tunable
degree of quantum fluctuations of the local moment
system \cite{Custers12,Aronson,Canfield,Lohneysen,Khalyavin}.
The global phase diagram accomodates a
Kondo-destroyed interacting quantum criticality, which corresponds to a Kondo destruction
at the AF quantum critical point \cite{Si-Nature,Coleman-JPCM,Paschen,Shishido2005}.
The Kondo destruction introduces new critical modes, and the critical theory goes beyond
a field theory for the fluctuations of the AF order parameter \cite{Hertz,Millis,Moriya}.

In order to access the variety of phases and transitions, it is important to treat the local moment antiferromagnetism
and Kondo singlet formation on an equal footing \cite{SiPaschen}.
Traditionally, the local moment AF phase and the heavy Fermi liquid
phase with static Kondo screening are more conveniently approached by different theoretical techniques.
The Kondo screening is more conveniently treated in a fermionic basis, as it yields fermionic quasiparticles
\cite{Hewson}. In particular,
within a large N analysis, the SU(2) symmetry of the local moment's spin operator $\mathbf{s}_i$ is enlarged
to SU(N) [or SP(N)], and the spin operator is written in terms of the slave fermions or spinons
$f_{i,\alpha}$, as $\mathbf{s}_i=f^{\dagger}_{i,\alpha} \boldsymbol \tau f_{i,\beta}$, where $\alpha=1,2,..,N$
and $\boldsymbol \tau$ are the generators of the SU(N) group. At the saddle point level, the static Kondo screening
is captured by the expectation value of the inter-species bilinear or hybridization operator
$\mathcal{O}_h=c^{\dagger}_{i,\alpha}f_{i,\alpha}+h.c$, where $c_{i,\alpha}$ is the creation operator
of the conduction fermions. This is the simplest local hybridization operator. However, we can also
consider a general non-local hybridization operator. In order to implement the constraint of half-filling
for the slave fermions one naturally introduces auxiliary gauge fields, which strongly interact with the $f$ fermions.
In such a description, the matter fields appear in the fundamental representation of the gauge group.
A nonzero expectation value of $\mathcal{O}_h$ corresponds to the Higgs phenomenon of the auxiliary gauge fields
and violates the local gauge invariance, which is disallowed by the Elitzur's theorem \cite{Elitzur}.
The gauge fluctuations
about the large N saddle point are important for restoring the local gauge invariance.
At least in the large $N$ limit,
the fluctuations  do not destroy the amplitude of the hybridization,
and heavy Fermi liquid can survive as a stable phase.
As there is no sharp distinction in the gauge sector between the Higgs phase
in the fundamental representation
and a confined phase \cite{FradkinShenker}, the compactness of the gauge field is not a severe issue
inside the heavy Fermi liquid phase.
The slave fermion method can also provide a qualitative description of a spin liquid
phase \cite{Burdin,Senthiletal,Pauletal},
which has a (vanishing $\langle \mathcal{O}_h \rangle$) Kondo destruction effect
\cite{Si-Nature,Coleman-JPCM},
and one rather considers the expectation values
of the various gauge symmetry breaking bilinears involving only the spinons.
However, in order to show that the spin liquid is a stable phase,
one has to prove that the matter-gauge field theory is in a deconfined phase.
In addition, the AF ordering is not captured on an equal footing,
because it can only be formulated for the SU(2) problem.

In contrast, magnetic order is more conveniently described in a bosonic basis. Indeed, a local order parameter
description works quite well inside the Kondo destroyed AF state, where the itinerant fermions move innocuously
in the periodic background potential provided by the AF order parameter. Depending on the presence
or absence of the commensuration between the band structure of the itinerant fermions and the periodic potential
generated by the AF order parameter, an insulating or a metallic state is realized, which is devoid of any static
Kondo screening.
In the absence of commensuration and in spatial dimensions higher than one, the stability of the AF metallic phase
has been addressed by employing a perturbative renormalization group analysis\cite{Yamamoto2007}.
When the conduction electrons are not integrated out, the Kondo coupling turns out to be a marginal operator,
and the AF metallic phase remains stable \cite{Yamamoto2007}.
Phase transitions can be considered in the renormalization-group procedure, after the conduction electrons are
integrated out \cite{Jones};
however, such a perturbative treatment in terms of the order parameter alone cannot
characterize the possible magnetically disordered states. In particular, the question arises
as to how the static Kondo screening can be captured in this basis.

In terms of the global phase diagram, the question is how
to use the non-linear-sigmal model basis to describe
the $P_L$ phase, the Kondo-screened state with a large Fermi surface, and the $P_S$ phase,
the Kondo-destroyed paramagnetic state with a small Fermi surface. The competition between
these two type of phases is illustrated by the red-dashed line in Fig.~\ref{Fig1}.

It is natural to expect that, to properly capture the magnetically-disordered side of the phase diagram,
we need to consider the non-perturbative topological defects of the order parameter field.
Within a coarse grained order parameter description of the local moments, the Berry's phase is the only quantity
that captures the quantized value of the spin, and plays a crucial role in determining the nature of the disordered
phase \cite{Haldane,Duncan,Read} and the
deconfined quantum critical point \cite{Senthiletal2} of the local moments.
The Berry's phase is normally tied with the instanton
configurations of the AF order parameter, and consequently we need to incorporate the scattering of the fermions
from these topological defects, for addressing the nature of the emergent magnetically disordered phase.

The fermion-instanton scattering has been addressed for the Kondo-Heisenberg model in
one dimension\cite{Tsvelik1994,Goswami1}. When the fermions are at half-filling and there is commensuration
between the band structure and the spin chain, the algebraic spin liquid phase of the spin half chain
is immediately destroyed in favor of a Kondo insulating state. It has been shown that the Berry's phase
of the spin half chain is canceled by an emergent Berry's phase due to the fermion-instanton scattering,
which leads to the spin gap\cite{Tsvelik1994,Goswami1}. Most importantly the issue of the Kondo
screening can be addressed
by using a composite order parameter $\langle \mathbf{n}_s \cdot \mathbf{n}_{\tau} \rangle$,
formed out of the
staggered magnetization operators of the local moments and the conduction
fermions\cite{Goswami1,Zachar,SiPivovarov}.
In Ref.~\onlinecite{Goswami1}, we have demonstrated the cancelation
of the Berry's phase and the emergence of a spin gap
even in the absence of the commensuration. Since, the Berry's phase of the spin chain is tied
with the fluctuating spin Peierls order, the cancelation of this geometric phase has been construed as a signature
of the competition between the Kondo screening and the spin-Peierls order \cite{Goswami1}.
This picture has also been established via bosonization based analysis.

In spite of the progresses, the connection between the composite order parameter
$\langle \mathbf{n}_s \cdot \mathbf{n}_{\tau} \rangle$
and a conventional description in terms of $\mathcal{O}_h$ has not been established yet.
In addition, the relevance of the topological defects in the Kondo singlet formation for higher spatial dimensions
has not been proven either.

In this paper we address these important issues. The scattering of fermions from topological defects is a venerable
problem of quantum field theory \cite{Rajaraman}, and little is known about its consequences for a generic
band structure of the fermions. However, considerable analytical progress can be made for the fermions
with linear dispersion. For this reason, we consider the two dimensional Kondo-Heisenberg model
on a honeycomb lattice at half-filling.
In this case, simple and concrete calculations can be preformed. Due to the choice of half-filling,
we restrict ourselves to the commensurate case,
and consequently deal with insulating states.

\subsection{Kondo singlet as topological defects of N\'eel order}

Describing the local moments in terms of a coarse grained quantum non-linear sigma model,
we will show that the
skyrmion defects of the antiferromagnetic order parameter host various competing order parameters.
More precisely we show that the skyrmion number is a conjugate variable of the competing orders.
For the Kondo lattice model,
we identify for the first time
Kondo singlet bilinears as the competing orders of the antiferromagnetism.
Some of these Kondo singlet bilinears are local, while some
break discrete symmetries of the lattice, and still some describe nonlocal second neighbor
Kondo hybridizations.

Inside the magnetically ordered phase, the skyrmion defects are finite energy topological excitations.
But, the skyrmion number is not changed by the tunneling events, as the hedgehogs and the antihedgehogs
remain linearly confined. Consequently, the competing orders, which are the conjugate variables
of the skyrmion number, remain as fluctuating quantities without acquiring expectation values.
On the paramagnetic side, the skyrmion number suffers strong quantum fluctuations
due to the tunneling events of the sigma model field, and the nucleation of the competing order
becomes possible. We identify a subset of the Kondo hybridizations, which can appear as the mass terms
for the underlying Dirac fermions, and within a weak coupling argument this subset becomes energetically favorable.
We also show that the antiferromagnetism and various competing singlet orders can be related to each other
via generalized chiral transformations of the underlying fermions.

\subsection{Competition between Kondo singlet and spin peierls phases}

We will also show that the conduction electrons acquire a Berry phase through their coupling to the hedgehog
configurations of the N\'eel order. Furthermore, this emergent Berry phase cancels that of the local moments.
Combined with the considerations of the competing orders which arise from the skyrmion defects,
our results demonstrate the competition between the Kondo-singlet formation and spin-Peierls order
when the antiferromagnetic order is suppressed.
We show that the difference between the Berry phases of the two subsystem is related to the possible competing orders.

The Kondo singlet phase supports Kondo resonance excitations, which are incorporated into the Fermi surface.
The resulting large Fermi surface is a defining property of the heavy-fermion state in the metallic case.
This corresponds
to the $P_L$ phase in the global phase diagram, Fig.~\ref{Fig1}. For the
commensurate filling we consider, a heavy-fermion band is fully filled and the chemical potential lies in the middle
of the Kondo hybridization gap; $P_L$ will therefore correspond to a Kondo insulator phase.

In the spin-Peierls phase, the static Kondo screening is destroyed. The Fermi surface will therefore
be entirely determined by the
conduction electrons. This corresponds to the $P_S$ phase in the phase diagram of Fig.~\ref{Fig1}.
For the commensurate problem, the Peierls order parameter gaps out the conduction fermions
and even the $P_S$ phase becomes an insulator. The Berry phase considerations suggest
the possibility of
exotic
non-Landau phase transition
 between the $AF_S$ and the $P_S$ phase,
as well as between the $AF_S$ and the $P_L$ phases,
for a class of discrete symmetry breaking Kondo hybridizations.

The remainder of the paper is organized as follows. In Sec.~\ref{model}, we describe the Kondo-Heisenberg
model on the
honeycomb lattice at half filling, and its continuum limit. In Sec.~\ref{topological_defects}, we discuss the topological
defects of the AF order. In Sec.~\ref{skymion_competing_orders}, we outline the procedure to construct the competing
orders of the skymion defects of the AF order in a fermionic representation. In Sec.~\ref{competing_order_Kondo},
we show
how the Kondo hybridization $\mathcal{O}_h$ arises from the core of the skymion defects of the AF order.
In Sec.~\ref{energetics}, we identify those Kondo hybridizations, which can appear as the Dirac masses and
are energetically favorable within weak coupling arguments. In Sec.~\ref{chiral rotation} we show that
the AF order parameter and the Kondo hybridizations are connected to each other via chiral transformations.
In Sec.~\ref{Berry_phase_emergent}, we demonstrate that conduction electrons moving in the topological backgrounds
of the AForder acquires a Berry phase. This emergent Berry phase cancels that of the local moments,
providing the most
explicit demonstration of the competition between the Kondo singlet state and the spin-peierls order.
Finally, in Sec.~\ref{conclusion}, we summarize our results and discuss the possible generalizations of our analysis.

\section{Kondo lattice model and its continuum limit}
\label{model}

\begin{figure}[htbp]
\includegraphics[scale=0.8]{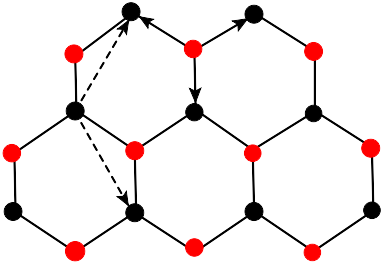}
\caption{(Color online)The red and the black circles respectively denote two interpenetrating triangular
sublattices of the honeycomb lattice. Two basis vectors of the triangular sublattice are shown
as the dashed lines with arrows. Three solid lines with arrows correspond to the nearest neighbor vectors.}
\label{Fig2}
\end{figure}

We focus on the following Kondo Heisenberg model on the honeycomb lattice at half filling
\begin{widetext}
\begin{eqnarray}
H_2=\sum_{\mathbf{r}_i \in A} \sum^3_{j=1} \bigg[-t_c \; c^{\dagger}_{A,\alpha}(\mathbf{r}_i)
c_{B,\alpha}(\mathbf{r}_i+\boldsymbol \delta_j)+h.c. +J_H \; \mathbf{s}_{A}(\mathbf{r}_i)
\cdot \mathbf{s}_{B}(\mathbf{r}_i+\boldsymbol \delta_j) +\frac{J_{K}}{2} c_{A,\alpha}^{\dagger}(\mathbf{r}_i) \;
\boldsymbol \sigma_{\alpha \beta} \; c_{A, \beta} (\mathbf{r}_i)\cdot \mathbf{s}_{A}(\mathbf{r}_i) \nonumber \\
+\frac{J_{K}}{6} c_{B,\alpha}^{\dagger}(\mathbf{r}_i+\boldsymbol \delta_j) \; \boldsymbol \sigma_{\alpha \beta} \;
c_{B, \beta} (\mathbf{r}_i+\boldsymbol \delta_j)\cdot \mathbf{s}_{B}(\mathbf{r}_i+\boldsymbol \delta_j)\bigg],
\end{eqnarray}
\end{widetext}
where $A$, $B$ denote two interpenetrating triangular sublattices, as shown in Fig.~\ref{Fig2}. The Pauli matrices
$\boldsymbol \sigma$ operate on the spin indices $\alpha$, $\beta$. The nearest neighbor hopping strength
is $t_c$ and $\boldsymbol \delta_j$ are the coordination vectors, which connect two sublattices,
and are shown in the Fig.~\ref{Fig2} as the solid lines with arrows. The explicit forms of the coordination vectors are
\begin{eqnarray}
\boldsymbol \delta_1=(0,-1)a , \: \: \boldsymbol \delta_2=(\sqrt{3},1)a/2, \: \: \boldsymbol \delta_3
=(-\sqrt{3},1)a/2.
\end{eqnarray}The local moments on the two sublattices are respectively represented by
$\mathbf{s}_{A}(\mathbf{r}_i)$ and $\mathbf{s}_{B}(\mathbf{r}_i+\boldsymbol \delta_j)$,
and $J_H$ is the nearest neighbor, AF Heisenberg coupling. The Kondo coupling is denoted
by $J_K$. We will consider both $J_H$ and $J_K$ to be antiferromagnetic (i.e., $>0$).

In the absence of the Kondo coupling, the fermion band structure is obtained by diagonalizing
the following Hamiltonian in the momentum space
\begin{eqnarray}
H_f=\sum_{\mathbf{k}}(c^\dagger_{A,\mathbf{k}}, c^\dagger_{B,\mathbf{k}} ) \left (F_{1,\mathbf{k}}\tau_1
+ F_{2,\mathbf{k}}\tau_2\right) \left(\begin{array}{c}
c_{A,\mathbf{k}} \\
c_{B,\mathbf{k}}\end{array} \right ),
\end{eqnarray}
where
\begin{eqnarray}
F_{1,\mathbf{k}}=\sum_j \cos \mathbf{k} \cdot \boldsymbol \delta_j, \: \:
F_{2,\mathbf{k}}=\sum_j \sin \mathbf{k} \cdot \boldsymbol \delta_j.
\end{eqnarray}
The energy spectrum of the two bands are given by
\begin{equation}
E_{n,\mathbf{k}}=(-1)^n \sqrt{F^2_{1,\mathbf{k}}+F^2_{2,\mathbf{k}}},
\end{equation}
where $n=1,2$ respectively correspond to the conduction and the valence bands. These two bands touch
at the six corners of the hexagonal Brillouin zone, and only two of them are inequivalent.
The spectrum can be linearized around these two inequivalent nodes, which we choose to be located at
\begin{equation}
\mathbf{K}_{\pm}=\pm \left(\frac{4 \pi}{3\sqrt{3}a},0\right).
\end{equation}

After linearizing the dispersion around these nodal points, the low energy quasiparticles are described
by the following real time action
\begin{eqnarray}
\mathcal{S}_f=\int d^2x dt \: \bar{\psi}_{\alpha}\bigg[i\gamma_{0}\otimes \sigma_0 \partial_{t}
+i v_\psi \gamma_j \otimes \sigma_{0}\partial_j \bigg]\psi_{\alpha},
\end{eqnarray}
where $v_\psi=\sqrt{3}t_c a/2$ is the Fermi velocity and the four component spinor is defined according to
$\psi_{\alpha}^{T}=(c_{A,+,\alpha},c_{B,+,\alpha},c_{B,-,\alpha}, c_{A,-,\alpha})$, with $\alpha$ being the spin index.
The fermion doublers at two inequivalent nodes are denoted by $\pm $ and the sublattice indices correspond to $A/B$.
The four component gamma matrices satisfy $\{ \gamma_{\mu}, \gamma_{\nu} \}=2 g_{\mu \nu}$,
with the metric $g_{\mu \nu}=(1,-1,-1)$, and $\bar{\psi}_{\alpha}=\psi^{\dagger}\gamma_0$. For convenience,
we work with the following chiral representation of the gamma matrices
\begin{equation}
\gamma_{0} =
 \left(\begin{array}{c c}
0 & \eta_0 \\
\eta_0 & 0
\end{array}\right), \;
\gamma_{j} =
 \left(\begin{array}{c c}
0 &  \eta_j \\
- \eta_j & 0
\end{array}\right), \;
\gamma_{5} =
 \left(\begin{array}{c c}
\eta_0 & 0\\
 0 & -\eta_0
\end{array}\right),
\end{equation}
where the Pauli matrices $\boldsymbol \eta$ operate on the sublattice indices. The free fermion action remains invariant under the following symmetry operations: (i) the inversion ($\mathcal{P}$), (ii) the reflections about the x ($\mathcal{I}_x$) and the y ($\mathcal{I}_y$) axes, and (iii) the time reversal transformation ($\mathcal{T}$). In order to transit to the Euclidean space,
we define the imaginary time $t \to -i \tau$, and $\gamma_j \to i\gamma_j$, and the Euclidean action becomes
\begin{eqnarray}
\mathcal{S}=\int d^2x d\tau \: \bar{\psi}_{\alpha}\bigg[\gamma_{0}\otimes \sigma_0 \partial_{\tau}
+ v_\psi \gamma_j \otimes \sigma_{0}\partial_j \bigg]\psi_{\alpha},
\end{eqnarray}

In the continuum limit the local moments are described by the following Euclidean
QNL$\sigma$M action\cite{Duncan,Chakravarty}
\begin{eqnarray}
\mathcal{S}_n=\frac{1}{2 c g}\int d^2x d\tau \left[c^2 (\partial_x \mathbf{n})^2+(\partial_{\tau}\mathbf{n})^2\right]
+ iS_B[\mathbf{n}].
\end{eqnarray}
The coupling constant $g$ has the dimension of length, and there is an antiferromagnetically ordered phase for
$g$ smaller than a critical strength $g_c$. The imaginary term $S_B[\mathbf{n}]$ corresponds to the Berry phase,
and this is an oscillatory quantity and does not have the continuum limit.

The Kondo coupling gives rise to the following scattering term between the conduction-electron
spinor $\psi$  and the QNL$\sigma$M field $\mathbf{n}$
\begin{equation}
\mathcal{S}_{fn}=g_K \int d^2x dt \; \bar{\psi}_{\alpha} \gamma_3 \mathbf{n} \cdot \boldsymbol
\sigma_{\alpha \beta} \psi_{\beta}.
\end{equation}
Inside the magnetically ordered state $\langle \mathbf{n} \rangle \neq 0$, and the Kondo coupling
acts as a mass term for the conduction fermions. This leads to an AF insulating state,
which is the analog of the Kondo destroyed AF$_S$ phase shown in Fig.~\ref{Fig1}. In addition, the staggered magnetization
of the conduction fermion is anti-aligned with $\mathbf{n}$. The magnetically ordered insulating phase remains stable
up to a critical ratio of the microscopic couplings $J_K/J_H$. Within the continuum description,
the enhancement of  $J_K/J_H$ increases the coupling $g$ for the QNL$\sigma$M, and eventually
destabilizes the AF phase. Inside the disordered phase, the Kondo coupling term is still a relevant perturbation,
and we expect $\langle \mathbf{n}_s \cdot \mathbf{n}_\tau \rangle =-1$ holds without having expectation values
of the independent staggered magnetizations. Therefore, we assume that the magnetic disordering occurs without
destroying the charge gap, which is inherited from the existence of the magnetization amplitude,
but no long range order (without phase stiffness). In such a situation, a coarse grained sigma model description
remains valid inside the magnetically disordered phase at the scale of the AF correlation length.
Given the commensuration between the fermion's band structure and the AF background, the existence
of the charge gap on either side of the transition is quite natural.

\section{Topological defects of the quantum non-linear sigma model
in 2+1 dimensions}
\label{topological_defects}

The sigma model in 2+1-dimensions have two important topological defects. There are static nonsingular topological
defects called skyrmions, which cost finite energy. A single skyrmion configuration satisfies the boundary
conditions $\mathbf{n}(r \to \infty)=\mathbf{n}^0$, where $r=\sqrt{x^2+y^2}$ and $\mathbf{n}^0$ is a constant
unit vector. Consequently the two dimensional space is compactified onto a two sphere $S^2$,
and the skyrmion configurations are described by the homotopy classification $\Pi_2(S^2)=\mathbb{Z}$.
The explicit form of the single skyrmion configuration and its topological index are respectively given by
\begin{eqnarray}
&&\mathbf{n}=\bigg(\frac{2r^q \lambda^q}{r^{2q}+\lambda^{2q}}\cos q\phi, \; \frac{2r^q \lambda^q}{r^{2q}
+\lambda^{2q}}\sin q\phi, \; \frac{r^{2q}-\lambda^{2q}}{r^{2q}+\lambda^{2q}}\bigg), \nonumber \\ \\
&&q_s=\frac{1}{8\pi} \int d^2x \; \epsilon_{\alpha \beta \lambda} \; \epsilon_{ij} \; n_{\alpha} \partial_i n_{\beta}
\partial_j n_{\lambda}=q,
\end{eqnarray}
where $\phi=\arctan (y/x)$.

In addition, we have the singular hedgehog configurations in the Euclidean space-time,
which change the skyrmion number of the background via tunneling and cost finite
action.\cite{Duncan,Read,MurthySachdev} These singular defects are also classified
according to $\Pi_2(S^2)=\mathbb{Z}$, but this involves the mapping of a sphere surrounding the
singularity onto another sphere. The corresponding topological invariant is given by
\begin{equation}\label{hedgehog}
q_h=\frac{1}{8\pi} \int d^2S_{a} \epsilon_{abc} \; \epsilon_{\alpha \beta \lambda} \;
n_{\alpha} \partial_b n_{\beta} \partial_c n_{\lambda},
\end{equation}
where the integral is performed on a sphere surrounding the hedgehog \cite{Arafune}.
The $q_h= \pm 1$ radial (anti)hedgehog corresponds to $\pm x_{\mu}/x$.

\begin{figure}[htbp]
\includegraphics[scale=0.6]{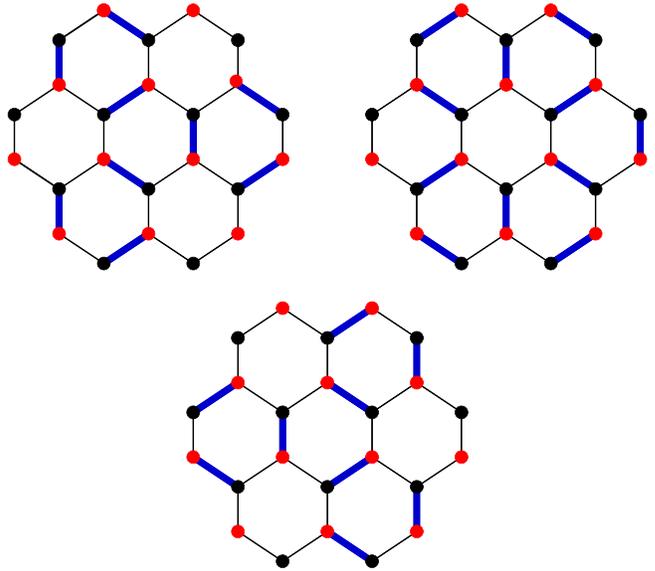}
\caption{(Color online) Three degenerate dimerization patterns for the spin Peierls order.
The dimerized bonds are marked as thick blue lines.}
\label{Fig3}
\end{figure}

The Berry's phase is related to the presence of these singular configurations. In the ordered phase,
the hedgehog and the anti-hedgehog are linearly confined, and the Berry's phase vanishes.
On the magnetically disordered side consideration of both the skyrmion and the hedgehogs become important.
In contrast to the (1+1) dimensions we do not have a continuum description for $S_B[\mathbf{n}]$.
In the paramagnetic phase the Berry's phase depends on the size of the spin $S$ and the lattice
coordination number $Z$ according to the formula\cite{Duncan,Read}
\begin{equation}\label{berry}
S_B[\mathbf{n}]=\int d\tau \sum_j \frac{4S \pi}{Z} \xi_j q_{h,j},
\end{equation}
where $j$ specifies the dual lattice sites, and $q_{h,j}$ is the topological charge of the hedgehogs located at $j$.
The dual lattice is partitioned into $Z$ sublattices and the integer valued weight factors are given by $\xi_j=0,1,...,Z-1$
on the different sublattices. Consequently, there is a periodicity $2S (\mathrm{modulo} \; Z)$.
On a honeycomb lattice $Z=3$, and the Berry's phase determines the pattern of the $C_{3v}$ symmetry breaking
due to the spin Peierls order for different quantized value of the spin. For $2S=0 (\mathrm{modulo} \; 3)$,
Berry's phase is absent and there is no spin Peierls order, and the disordered ground state is nondegenerate.
When $2S=1 (\mathrm{modulo} \; 3)$, the disordered ground state has threefold degeneracy,
and corresponds to the Peierls order, as shown in Fig.~\ref{Fig3}.

\section{Skyrmions and competing orders in the fermionic representation}
\label{skymion_competing_orders}

Due to the absence of a continuum representation of Berry's phase above one dimension, it is harder to analyze
its effects within the coarse grained representation. However, we can understand some of the competing orders
present in the core of topological defects by introducing auxiliary fermions for describing
the local moments \cite{AffleckHaldane, Hermele, FisherSenthil,TanakaHu}. For simplicity we can assume
that the auxiliary fermions only hop to the nearest neighbor sites like the conduction fermions,
with a hopping strength $t_f$. At low energy these fermions can be also described by the Dirac equation
for a new set of spinor $\chi$. The auxiliary fermions interact via a strong Hubbard $U$,
which leads to the AF ordering above a threshold value $U_c$. Inside the AF phase,
when we freeze the amplitude of the order parameter, the fermion-boson coupling takes the form
\begin{eqnarray}\label{chi}
\mathcal{S}_{\chi}=\int d^2x dt \: \bar{\chi}_{\alpha}\bigg [i\gamma_{0}\otimes \sigma_0 \partial_{\tau}
+ v_\chi i \gamma_j \otimes \sigma_{0}\partial_j \nonumber \\
+ g_{\chi} \gamma_3 \mathbf{n}\cdot \boldsymbol \sigma \bigg]_{\alpha \: \beta}\chi_{\beta},
\end{eqnarray}
where $v_{\chi}=\sqrt{3}t_f a/2$. Due to the anti-alignment of the staggered magnetizations of the $\psi$
and the $\chi$ fermions, the product $g_{\psi} g_{\chi} <0$.

The competition between the AF and the Peierls phase of the local moments can be understood in the following way.
The Peierls order parameter shown in Fig.~\ref{Fig3} can be represented as the Kekule bond-density wave order
of the $\chi$
fermions \cite{Chamon1}. In the presence of this bond density order the hopping strengths are modified
in the following way
\begin{eqnarray}
H_f=-\sum_{\mathbf{r}_i \in A} \; \sum_{j} \; \left(t_f+\delta t_{i,j}\right)f^{\dagger}_{A}(\mathbf{r}_i)f_{B}
(\mathbf{r}_i+\boldsymbol \delta_j)+h.c. ,\nonumber \\ \end{eqnarray}
where
\begin{equation}
\delta t_{i,j}= m \exp \left[i \mathbf{K}\cdot \left(\boldsymbol \delta_j + 2 \mathbf{r}_i \right)+i \phi \right].
\end{equation}
This density wave has a $\mathbf{Q}=\mathbf{K}_+-\mathbf{K}_-=2\mathbf{K}_+$ modulation,
and couples the two opposite sublattices and the inequivalent valleys simultaneously.
The fluctuations of this order is captured by the spatial dependence of the amplitude $m$ and the phase $\phi$.
In the continuum limit, the Kekule order couples to the following inter-valley and inter-sublattice bilinears
\begin{equation}\label{Ok}
\mathcal{O}_K=m\; \left(\cos \phi \bar{\chi}\chi+i \sin \phi \bar{\chi}\gamma_5\chi\right)
=m \; \bar{\chi} e^{i \phi \gamma_5}\chi.
\end{equation}Thus the matrix $\gamma_5$ is the generator of the $U(1)$ chiral transformation
between the two Kekule bilinears \cite{Jackiw}. These two bilinears anticommute with the kinetic energy
and act as the Dirac mass.

The competition among the singlet and triplet orders can be illustrated by evaluating the induced
chiral current in the presence of the skyrmion defects of the $\mathbf{n}$ field \cite{Wilczek1},
which we pursue now.
\begin{figure}[htbp]
\includegraphics[width=5cm,height=4cm]{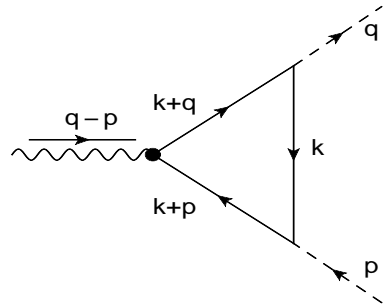}
\caption{(Color online)The pertinent triangle diagram for calculating the induced chiral current
in the presence of the skyrmion defects of the magnetic order parameter. The solid and the dashed lines respectively
denote the fermionic and the QNL$\sigma$M fields. The wavy line describes the fictitious chiral gauge field,
and the chiral vertex is denoted by the solid black circle.}
\label{Fig4}
\end{figure}
Since $\gamma_5$ is the generator of the chiral rotations, the nucleation of the Kekule order
causes the Meissner effect of the fictitious chiral gauge field $a_{\mu}$ \cite{Jackiw}.
This gauge field couples minimally to the chiral current as $\bar{\chi}_{\alpha} \gamma_5 \gamma_{\mu} a_{\mu}
\chi_{\alpha}$, and the gauged action is
\begin{eqnarray}
\mathcal{S}_{\chi,1}&=&\int d^2x dt \: \bar{\chi}_{\alpha}\bigg [i\gamma_{0}\otimes \sigma_0 (\partial_{t}
+i a_t \gamma_5)+ v_\chi \gamma_j \otimes \sigma_{0} \nonumber \\ &&\times (\partial_j +i a_j \gamma_5)
-m e^{i \phi \gamma_5}\bigg]\chi_{\alpha}
\end{eqnarray}
Under the chiral transformation
\begin{equation}
\chi \to e^{-i \alpha/2 \gamma_5} \chi, \: \: \bar{\chi} \to \bar{\chi} e^{-i \alpha/2 \gamma_5}
\end{equation}
the $\phi \to \phi-\alpha$ and $a_\mu \to a_\mu- \partial_\mu \alpha /2$. Inside the Neel ordered phase,
there is no expectation value for the Peierls order. But, we can assess its role as a fluctuating order
by computing the expectation value of the chiral current inside the AF phase from the following action
\begin{eqnarray}\label{chi}
\mathcal{S}_{\chi,2}&=&\int d^2x dt \: \bar{\chi}_{\alpha}\bigg [i\gamma_{0}\otimes \sigma_0 (\partial_{t}
+i a_t \gamma_5)+ v_\chi \gamma_j \otimes \sigma_{0} \nonumber \\ &&\times (\partial_j +i a_j \gamma_5)
+ g_{\chi}  \gamma_3 \mathbf{n}\cdot \boldsymbol \sigma \bigg]_{\alpha \: \beta}\chi_{\beta}
\end{eqnarray} The expectation value of the chiral current is determined as $j_{ch,\chi,\mu}
= \delta S_{\chi,2} /\delta a_{\mu}|_{a_\mu=0}$.

The chiral current $j_{ch,\chi,\mu}$ receives a nontrivial contribution from the skyrmion configurations
of the $\mathbf{n}$ field \cite{Tanaka1, FuSachdev}. This can be seen in the process of gradient expansion
by considering the triangle diagram in Fig. 3.  The explicit expression is obtained in the following manner
\begin{eqnarray}
&&\langle \bar{\chi}_{\alpha} \gamma_5 \gamma_{\mu} \chi_{\alpha} \rangle=\mathrm{Tr}
\bigg[\frac{\gamma_{\mu}\gamma_5}{\gamma_{\mu}\partial_{\mu}+  g_{\chi} \gamma_3 \mathbf{n}
\cdot \boldsymbol \sigma}\bigg] \nonumber \\
&&=\mathrm{Tr}\bigg[\frac{\gamma_{\mu}\gamma_5\left(\gamma_{\nu}\partial_{\nu}+  g_{\chi} \gamma_3
\mathbf{n}\cdot \boldsymbol \sigma \right)}{\partial^2+g^2_\chi+g_{\chi} \gamma_{\rho} \gamma_3 \partial_{\rho}
\mathbf{n} \cdot \boldsymbol \sigma}\bigg] \nonumber \\
&&=g^3_{\chi}\mathrm{Tr}\bigg[\frac{\gamma_{\mu}\gamma_5 \gamma_{3} \gamma_{\nu} \gamma_{3}
\gamma_{\lambda} \gamma_3 \mathbf{n}\cdot \boldsymbol \sigma  \partial_{\nu}\mathbf{n}\cdot \boldsymbol
\sigma \partial_{\lambda}\mathbf{n}\cdot \boldsymbol \sigma}{\left(\partial^2+g^2_{\chi}\right)^3}\bigg]
\end{eqnarray}
The trace in the above formula consists of a matrix trace and also integral over the spatial coordinates.
The matrix trace leads to $8 \times \epsilon_{\mu \nu \lambda} \times \epsilon_{abc}$,
and after using the following elementary integral in the energy-momentum space
\begin{equation}
\int \frac{d^3k}{(2\pi)^3} \frac{1}{(k^2+g^2_{\chi})^3}=\frac{16 \pi}{|g_{\chi}|^3},
\end{equation}
we obtain
\begin{eqnarray}\label{sk1}
j_{ch, \chi, \mu}=\langle \bar{\chi}_{\alpha} \gamma_5 \gamma_{\mu} \chi_{\alpha} \rangle
= \frac{\mathrm{sgn}(g_\chi)}{4\pi}\epsilon_{\mu \nu \lambda} \mathbf{n}\cdot(\partial_{\nu}\mathbf{n}
\times \partial_{\lambda}\mathbf{n}) . \nonumber \\
\end{eqnarray}
Therefore, the chiral current density equals to the topological skyrmion current.

This result is in fact tied with the \emph{parity anomaly} of two component Dirac fermions in the following way.
In the action of the $\chi$ fermions, two valleys are decoupled. If we focus on a single valley with two component
Dirac fermions, the model breaks reflection/parity symmetry, which is the source of the parity anomaly
\cite{Deser,Semenoff,Redlich,Fradkin1}.
For our four component fermions, the expectation value of the standard electromagnetic current vanishes,
i.e, $\langle \bar{\chi} \gamma_\mu \chi \rangle=0$. If we add and subtract the electromagnetic
and the chiral currents
of the four component fermions, and subsequently divide by the factor of two, we arrive at the projected
electromagnetic current of the two component fermions at each valley
\begin{eqnarray}\label{sk2}
&&j_{em,\mu,\pm}=\langle \bar{\chi}_{\alpha}\frac{ (1 \pm \gamma_5)}{2} \gamma_{\mu} \chi_{\alpha} \rangle
\nonumber \\
&&=\pm \frac{\mathrm{sgn}(g_\chi)}{8\pi}\epsilon_{\mu \nu \lambda} \mathbf{n}\cdot(\partial_{\nu}\mathbf{n}
\times \partial_{\lambda}\mathbf{n}).
\end{eqnarray}
When we consider both valleys, the parity is restored, and the net electromagnetic current vanishes,
leaving a nonzero chiral current of Eq.~(\ref{sk1}). Notice that the chiral current equals to the skyrmion current,
in contrast to the two component problem, where the induced electromagnetic current is
one half of the skyrmion current\cite{Jaroszewicz,Wilczek2, Abanov}.

In the absence of the hedgehog events, the skyrmion and the chiral currents are conserved. This conservation
certainly holds inside the AF phase, as the hedgehogs and the antihedgehogs remain linearly confined inside
the magnetically ordered phase. The chiral charge $\mathcal{Q}$, which acts as the generator of the chiral
$U(1)$ rotational symmetry is given by the skyrmion number of the background
\begin{eqnarray}
&&\mathcal{Q}_{\chi}=\int d^2x \langle \bar{\chi}_{\alpha} \gamma_5 \gamma_{0} \chi_{\alpha} \rangle
= \frac{\mathrm{sgn}(g_\chi)}{4\pi}\int d^2x  \mathbf{n}\cdot(\partial_{1}\mathbf{n} \times \partial_{2}\mathbf{n})
\nonumber \\&&=\mathrm{sgn}(g_\chi) \; q
\end{eqnarray} In the presence of the hedgehogs, the conservation law breaks down.
If we consider a spherical region surrounding the hedgehog singularity, we can apply Gauss' divergence theorem
to obtain
\begin{eqnarray}
&&\int d^3x \partial_{\mu} j_{ch, \chi, \mu}
=\int dS j_{ch, \chi, \hat{r}}\nonumber \\&&=\frac{r^2}{4\pi} \int^\pi_0 \sin \theta d\theta \int^{2\pi}_0 d\phi \;
\epsilon_{r \theta \phi}\mathbf{n} \cdot \left(\partial_\theta \mathbf{n} \times \partial_\phi \mathbf{n}\right) =q_h
\nonumber \\
\end{eqnarray}
where $\epsilon_{r \theta \phi}=r^{-2}$. Therefore, inside the magnetically disordered phase we have
very strong quantum fluctuations of the $\mathcal{Q}_{\chi}$, and at each the singular tunneling event
the skyrmion number of the background jumps instantaneously by the hedgehog's topological charge.
For this reason, $\mathcal{Q}_{\chi}$ is a fast variable in the disordered phase,
and its conjugate will serve as the slow variable and appropriate competing order.

If we restrict ourselves to the singlet orders in the particle hole channel,
\begin{equation}\label{comp1}
\mathcal{O}_{M,\chi}=\bar{\chi} \hat{M} \otimes \sigma_0 \; e^{i \phi \gamma_5} \chi, \: \: [\hat{M}, \gamma_5]=0
\end{equation}
serve as the possible competing orders, where $\hat{M}$ is a $4\times 4$ matrix.
The commutation relation $ [\hat{M}, \gamma_5]=0$ indeed implies
\begin{equation}
\left[\mathcal{O}_{M,\chi},\mathcal{Q}_{\chi}\right]=2 i \bar{\chi} \hat{M} e^{i\left(\frac{\pi}{2}
+ \phi \right) \gamma_5} \chi
\end{equation}
and $\mathcal{Q}_{\chi}$ causes rotation between the two components of $\mathcal{O}_{M,\chi}$.
Thus, $\mathcal{O}_{M,\chi}$ are indeed the slow conjugate variables of the chiral charge. The following eight matrices
\begin{equation}
\hat{M}=1, \gamma_5, [\gamma_0,\gamma_j]/2i, [\gamma_i,\gamma_j]/2i
\end{equation}
where $i,j=0,1,2,3$, commute with $\gamma_5$. Not all of these choices are independent.
For example both $1$ and $\gamma_5$ correspond to the Peierls order. The explicit forms of the matrices
$\Sigma_{0j}=[\gamma_0,\gamma_j]/2i$, $\Sigma_{jk}=[\gamma_j,\gamma_k]/2i$ are respectively given by
\begin{equation}
\Sigma_{0j} =
 \left(\begin{array}{c c}
i \eta_j & 0 \\
0 & -i \eta_j
\end{array}\right), \;
\Sigma_{jk} =
i \epsilon_{ljk} \left(\begin{array}{c c}
\eta_l &  0 \\
0 & \eta_l
\end{array}\right).
\end{equation}
Therefore, $\Sigma_{0l}=\gamma_5 \epsilon_{ljk} \Sigma_{jk}$ and apart from Peierls order,
we have only three independent singlet competing order pairs, which can be obtained by using $\hat{M}=\Sigma_{0l}$.
All the bilinears can be physically described as appropriate charge or current density wave orders,
with modulation wavevector $2 \mathbf{K}_+$\cite{FuSachdev}. For $l=3$, we obtain the intervalley, intersublattice
current density wave order. Both $l=1,2$ describe intrasublattice charge density wave orders. Out of these two,
$l=2$ corresponds to a sublattice staggered charge density wave. Hence, the core of the skyrmion excitation
carries several fluctuating competing orders \cite{FisherSenthil,TanakaHu,Tanaka1,FuSachdev}.

Inside the AF phase the skyrmion excitations and accordingly all the competing order parameters are gapped.
Only in the disordered phase, the spin stiffness vanishes and it becomes viable to nucleate the competing order.
Among all the competing orders, only the spin Peierls bilinear anticommutes with the kinetic energy and the AF order,
and can maximize the gap in the fermion's spectrum. Therefore, within a weak coupling argument the spin Peierls order
will be favored over the other competing orders. The Berry's phase dictates the pattern in which the tunneling
singularities are arranged, and consequently determines the expectation value of the chiral angle $\phi$
for the emergent ground state. Now we apply this strategy of the chiral current computation in the presence
of the Kondo coupling for exposing the new competing ordered states.

\section{Competing orders in the presence of the Kondo coupling}
\label{competing_order_Kondo}

We begin from the AF insulating state where the staggered magnetizations of the $\psi$ and $\chi$ fermions
are anti-aligned. The effective action for both species can be compactly described by
\begin{eqnarray}\label{skondo1}
&&\mathcal{S}= \int d^2x dt \bar{\Psi} \bigg[ i\gamma_0 \otimes \sigma_0 \otimes\tau_0 \partial_0
+i v_+\gamma_j \otimes \sigma_0 \otimes \tau_0 \partial_j   \nonumber \\ &&+i v_-\gamma_j \otimes
\sigma_0 \otimes \tau_3 \partial_j + g_+ \gamma_3 \mathbf{n} \cdot \boldsymbol \sigma \tau_0
+ g_-\gamma_3 \mathbf{n} \cdot \boldsymbol \sigma \tau_3 \bigg]\Psi,\nonumber \\
\label{s1}
\end{eqnarray}
where $\Psi^T=(\psi^T, \chi^T)$, and Pauli matrices $\boldsymbol \tau$ operate on the species label,
and $v_{\pm}=(v_{\psi} \pm v_{\chi})/2$, $g_{\pm}=(g_{\psi} \pm g_{\chi})/2$. In addition to the $j_{ch, \chi, \mu}$
of Eq.~(\ref{sk1}), we also have to account for the chiral current of the $\psi$ fermions
\begin{eqnarray}\label{sk2}
j_{ch, \psi, \mu}=\langle \bar{\psi}_{\alpha} \gamma_5 \gamma_{\mu} \psi_{\alpha} \rangle= \frac{\mathrm{sgn}(g_\psi)}
{4\pi}\epsilon_{\mu \nu \lambda} \mathbf{n}\cdot(\partial_{\nu}\mathbf{n}
\times \partial_{\lambda}\mathbf{n}) . \nonumber \\
\end{eqnarray}

Now we take the sum and the difference between the chiral currents of two species, and respectively denote them as
\begin{eqnarray}
j_{T,ch,\mu}=\bar{\Psi}\gamma_{\mu}\gamma_5\otimes \sigma_0 \otimes \tau_0\Psi, \:
j_{d,ch, \mu}=\bar{\Psi}\gamma_{\mu}\gamma_5\otimes \sigma_0 \otimes \tau_3 \Psi. \nonumber \\
\end{eqnarray}
With the aid of Eq.~(\ref{sk1}) and Eq.~(\ref{sk2}), we find
\begin{eqnarray}\label{sum}
j_{T,ch,\mu}= \frac{[\mathrm{sgn}(g_\chi)+\mathrm{sgn}(g_\psi)]}{4\pi}\epsilon_{\mu \nu \lambda} \mathbf{n}
\cdot(\partial_{\nu}\mathbf{n} \times \partial_{\lambda}\mathbf{n})=0, \nonumber \\
\end{eqnarray}
and
\begin{eqnarray}\label{diff}
j_{d,ch,\mu}&=&\frac{[\mathrm{sgn}(g_\chi)-\mathrm{sgn}(g_\psi)]}{4\pi}\epsilon_{\mu \nu \lambda} \mathbf{n}
\cdot(\partial_{\nu}\mathbf{n} \times \partial_{\lambda}\mathbf{n}) \nonumber \\
&=&\frac{2\mathrm{sgn}(g_\chi)}{4\pi}\epsilon_{\mu \nu \lambda} \mathbf{n}\cdot(\partial_{\nu}\mathbf{n}
\times \partial_{\lambda}\mathbf{n}) ,
\end{eqnarray}
where we have used $g_{\psi}g_{\chi}<0$.

As the total chiral current $j_{T,ch,\mu}=0$, the skyrmion number does not generate the competing orders, which are conjugate variables of the total chiral charge $\bar{\Psi} \gamma_0 \gamma_5 \otimes \sigma_0 \otimes \tau_0 \Psi$.
Instead, the difference between two chiral currents $j_{d,ch,\mu}$ is equal to twice the skyrmion current,
and
\begin{eqnarray}
\mathcal{Q}_{\Psi,-}=\int d^2x \langle \bar{\Psi} \gamma_5 \gamma_{0} \sigma_0 \otimes \tau_3 \Psi \rangle
= 2 \mathrm{sgn}(g_\chi) \; q_s
\end{eqnarray}
will act as the generator of interesting competing orders.

In order to determine the appropriate competing
order parameters, we first add a fictitious gauge field $\mathcal{A}_{\mu}$ in Eq.~(\ref{skondo1}),
which couples to the current $j_{d,ch,\mu}$ to obtain
\begin{eqnarray}
&&\mathcal{S}= \int d^2x dt \bar{\Psi} \bigg[ i\gamma_0 \otimes \sigma_0 \otimes\tau_0 (\partial_0
+i \mathcal{A}_0 \gamma_5 \tau_3) \nonumber \\
&&+i v_+\gamma_j \otimes \sigma_0 \otimes \tau_0 (\partial_0 +i \mathcal{A}_j \gamma_5 \tau_3)
+i v_-\gamma_j \otimes \sigma_0 \otimes \tau_3 (\partial_0 \nonumber \\ && +i \mathcal{A}_j \gamma_5 \tau_3)
+ g_+ \gamma_3 \mathbf{n} \cdot \boldsymbol \sigma \tau_0 + g_-\gamma_3 \mathbf{n} \cdot \boldsymbol
\sigma \tau_3 \bigg]\Psi,
\label{s2}
\end{eqnarray}
We again focus on the spin singlet competing orders in the particle hole channel. The following bilinears
\begin{equation}\label{compkondo1}
\mathcal{O}_{\mathcal{M}}=\bar{\Psi} \hat{\mathcal{M}} \otimes \sigma_0 \exp \left( i \phi
\gamma_5 \tau_3 \right)  \Psi, \: \: [ \hat{\mathcal{M}}, \gamma_5 \tau_3]=0
\end{equation}
will be the appropriate candidate for the competing orders, where $\hat{\mathcal{M}}$ is a $8 \times 8$ matrix.
Under the chiral gauge transformations,
\begin{equation}\label{chdt}
\Psi \to \exp \left( i \frac{\phi}{2} \; \gamma_5 \tau_3 \right)  \Psi; \: \bar{\Psi} \to \bar{\Psi} \exp \left( i \frac{\phi}{2} \;
\gamma_5 \tau_3 \right),
\end{equation} the gauge potential $\mathcal{A}_{\mu} \to \mathcal{A}_{\mu} + \partial_{\mu} \phi /2 $,
and the nucleation of any of the competing orders described in Eq.~(\ref{compkondo1}) causes Meissner effect
of the $\mathcal{A}_{\mu}$. We also find that
\begin{equation}
\left[\mathcal{O}_{\mathcal{M}},\mathcal{Q}_{\Psi,-}\right]=2 i \bar{\Psi} \hat{\mathcal{M}}\otimes
\sigma_0 e^{i\left(\frac{\pi}{2}+ \phi \right) \gamma_5 \tau_3} \Psi,
\end{equation}
which further justifies the role of $\mathcal{O}_{\mathcal{M}}$ as the competing order.

The condition $[\mathcal{M}, \gamma_5 \tau_3 ]=0$ can be satisfied in the following ways:
\begin{eqnarray}
&&(i) \; \{\hat{\mathcal{M}}, \gamma_5 \}=0, \; \{\hat{\mathcal{M}}, \tau_3 \}=0, \label{mat1}\\
&&(ii) \; [\hat{\mathcal{M}},\gamma_5]=0, \; [\hat{\mathcal{M}},\tau_3]=0.\label{mat2}
\end{eqnarray}
The case (i) describes the inter-species Kondo singlet bilinears, and
\begin{eqnarray}
\hat{\mathcal{M}}=\gamma_\mu \otimes \tau_{1}, \: \gamma_\mu \gamma_5 \otimes \tau_{1}, \:
\gamma_\mu \otimes \tau_{2}, \: \gamma_\mu \gamma_5 \otimes \tau_{2}
\end{eqnarray}
where $\mu=0,1,2,3$. The choice of $\gamma_{\mu}$ and $\gamma_{\mu}\gamma_5$ are not independent,
and we can only focus on
$$\hat{\mathcal{M}}=\gamma_\mu \otimes \tau_{1/2}$$ for capturing the independent bilinears.
Note that, the choice of $\tau_{1}$ and $\tau_{2}$ are connected by a transformation
${\rm e}^{i\phi \tau_3}$, which accounts for the number difference of $\psi$ and $\chi$ fermions.
These do not represent independent operators, because the staggered $U(1)$ gauge symmetry
is already broken by the Kondo-singlet formation. In the remainder of this section, this same sense
applies whenever the combination $ \tau_{1/2}$ appears.
It is also
important to note that the Kondo singlet bilinears do not cause any valley mixing and do not carry
momentum $2 \mathbf{K}_+$.

In particular, $\hat{\mathcal{M}}=\gamma_0 \otimes \tau_{1/2}$ leads to
\begin{eqnarray}\label{kb1}
\mathcal{O}_{\mathcal{M}}&=&\bar{\Psi} \gamma_0 \left(\cos \phi \tau_{1/2} \pm \sin \phi \gamma_5
\tau_{2/1} \right)\Psi \\
&=&\Psi^{\dagger} \left(\cos \phi \tau_{1/2} \pm \sin \phi \gamma_5 \tau_{2/1} \right)\Psi.
\end{eqnarray}
Notice that the $\gamma_5$ independent parts $\Psi^\dagger \tau_{1/2} \Psi$ correspond to
the conventional onsite, Kondo singlet blilinears
\begin{eqnarray}
\Psi^\dagger \tau_{1/2} \Psi \equiv \left(c^\dagger_{A,i} \; c^\dagger_{B,i} \; f^\dagger_{A,i} \;
f^\dagger_{B,i}\right) \tau_{1,2} \otimes \eta_0
 \left(\begin{array}{c}
c_{A,i} \\
c_{B,i} \\ f_{A,i} \\
f_{B,i}
\end{array}\right)+ h.c. \nonumber \\
\end{eqnarray}
where $i$ denotes the label for the two atom unit cell of the honeycomb lattice. For the conventional Kondo
singlet bilinear, the sign of the hybridization remains the same for both valleys. In contrast, the $\gamma_5$
dependent parts have opposite amplitudes at the two valleys, thereby breaking the inversion symmetry
(however, it does not break time-reversal symmetry).
We will come back to its lattice version later.

When $\hat{\mathcal{M}}=\gamma_3 \otimes \tau_{1/2}$, we find
\begin{eqnarray}\label{kb2}
\mathcal{O}_{\mathcal{M}}&=&\bar{\Psi} \left(\cos \phi \gamma_3 \otimes \tau_{1,2} + \sin \phi \gamma_3
\gamma_5 \tau_{2,1} \right)\\
&=&\Psi^\dagger\left(\cos \phi \gamma_0\gamma_3 \otimes \tau_{1,2} + \sin \phi \gamma_0\gamma_3
\gamma_5 \tau_{2,1} \right)
\end{eqnarray}
The $\gamma_5$ independent parts
\begin{eqnarray}
\Psi^\dagger \gamma_0 \gamma_3 \tau_{1/2} \Psi & \equiv & \left(c^\dagger_{A,i} \;
c^\dagger_{B,i} \; f^\dagger_{A,i} \; f^\dagger_{B,i}\right) \tau_{1,2} \otimes \eta_3
 \left(\begin{array}{c}
c_{A,i} \\
c_{B,i} \\ f_{A,i} \\
f_{B,i}
\end{array}\right) \nonumber \\ &+& h.c.,
\end{eqnarray}
which describe sublattice staggered hybridizations, and do not change sign between two valleys.
These Kondo singlets break the inversion symmetry of the lattice. The $\gamma_5$ dependent parts change sign
between two valleys and break reflection symmetries with respect to the x and the y axes, and we again address them separately.

When $\hat{\mathcal{M}}=\gamma_{1/2} \otimes \tau_{1/2}$, we find
\begin{eqnarray}\label{kb3}
\mathcal{O}_{\mathcal{M}}&=&\bar{\Psi} \left(\cos \phi \gamma_{1/2} \otimes \tau_{1,2}
+ \sin \phi \gamma_{1/2} \gamma_5 \tau_{2,1} \right)\\
&=&\Psi^\dagger\left(\cos \phi \gamma_0\gamma_{1/2} \otimes \tau_{1,2} + \sin \phi \gamma_0\gamma_{1/2}
\gamma_5 \tau_{2,1} \right)
\end{eqnarray}The $\gamma_5$ independent parts
\begin{eqnarray}
&&\Psi^\dagger \gamma_0 \gamma_{1/2} \tau_{1/2} \Psi  \equiv  \nonumber \\
&&\left(c^\dagger_{A,i} \; c^\dagger_{B,i} \; f^\dagger_{A,i} \; f^\dagger_{B,i}\right) \tau_{1,2}
\otimes \eta_{1/2}
 \left(\begin{array}{c}
c_{A,i} \\
c_{B,i} \\ f_{A,i} \\
f_{B,i}
\end{array}\right) + h.c.,
\end{eqnarray} which describe inter-sublattice hybridizations, which do not change sign between two valleys. Both of these break inversion symmetry of the honeycomb lattice.

Now we briefly discuss the lattice realizations of the $\gamma_5$ dependent Kondo bilinears.
The criterion for obtaining a staggered order in the valley sector has been established by Haldane
in the context of anomalous quantum Hall state \cite{HaldaneHall}. For the anomalous charge quantum Hall state,
we have simultaneous staggering in the sublattice and the valley sectors. This is obtained from the following imaginary,
next nearest neighbor terms
$$i t^{\prime}  \sum_{<<ij>>} \nu_{ij} c^\dagger_{i}c_j +h.c.,$$
where $<<ij>>$ describe the next nearest neighbors, and the sublattice staggering is captured
by $\nu_{ij}=\pm 1$ respectively for A and B sublattices. This gives rise to a sublattice and valley staggered
Dirac mass for the continuum theory. If we choose the following nonlocal hybridization
\begin{equation}
i \sum_{<<ij>>} \left(c^\dagger_{A,i} \; c^\dagger_{B,i} \; f^\dagger_{A,i} \; f^\dagger_{B,i}\right) \tau_{1,2}
\otimes \eta_0
 \left(\begin{array}{c}
c_{A,j} \\
c_{B,j} \\ f_{A,j} \\
f_{B,j}
\end{array}\right)+h.c.,
\end{equation}
in the continuum limit we obtain $\Psi^\dagger \gamma_5 \tau_{1/2} \Psi$. If we introduce the sublattice staggering
factor $\nu_{ij}$ in the above formula, we obtain $\Psi^\dagger \gamma_0 \gamma_3 \gamma_5 \tau_{1/2}\Psi$.

The case (ii) in Eq.~(\ref{mat2}) describe the order parameters with valley mixing, which carry $2K_+$ momentum.
In addition these order parameters are diagonal in the species label. For simplicity, we only discuss one term
\begin{equation}
\hat{\mathcal{M}}=\mathbb{I} \otimes \tau_{0,3}
\end{equation}
which is related to the Peierls order. For this choice
\begin{equation}\label{stsp}
\mathcal{O}_{\mathcal{M}}=\bar{\Psi} \left(\cos \phi \tau_{0/3}+ i \sin \phi \gamma_5 \tau_{3/0} \right)\Psi
\end{equation}
which is a superposition of two opposite types of Peierls patterns for the two different species.

Therefore, the skyrmion core carries many fluctuating Kondo singlet, as well as
translational symmetry breaking orders.
At the level of the effective theory, we can not determine which $\mathcal{O}$ is realized in the disordered phase.
However, we can obtain some intuition regarding the energetics by considering some simple limiting cases,
as shown in the following section.

\section{Simple consideration of energetics}
\label{energetics}
For a single species of fermions, we have shown that only the spin Peierls order anticommutes
with the AF order and also with the kinetic energy. Consequently, only the combination of the AF
and Peierls order appear as a general mass term; nucleation of either of these orders gaps
out the Dirac points. Therefore, on the magnetically disordered side the Peierls order within simple weak coupling
arguments lead to maximal gain in the condensation energy, and appears to be the most favorable competing order.
 Following this line of reasoning, we can ask which competing orders can appear as the Dirac mass
 in the Kondo-lattice case with two species of fermions.

Since the Peierls order parameter of Eq.~(\ref{stsp}) does not mix two species,
it anticommutes with full Hamiltonian, and remains as a Dirac mass. For the arbitrary $v_\pm$ and $g_\pm$,
none of the Kondo singlet bilinears anticommute with the full Hamiltonian.
We will however consider two simple
limiting cases
$$(i) v_+=0, g_+=0; \: \: \: (ii) v_-=0, g_+=0.$$
We have chosen to set $g_+=0$, as $g_-$ is the dominant coupling
with AF order parameter.

For case (i), the Hamiltonian is given by
\begin{equation}\label{h1}
H_1=-i v_- \gamma_0 \gamma_j \otimes \sigma_0 \otimes \tau_3 \partial_j
+  g_- \gamma_0 \gamma_3 \mathbf{n} \cdot \boldsymbol \sigma \tau_3
\end{equation} Now the following Kondo singlet bilinears from Eq.~(\ref{kb1})
$ (i) \Psi^\dagger ( \cos \alpha \tau_1 +\sin \alpha \tau_2)\Psi$,
$(ii) \Psi^\dagger (\cos \phi \tau_1
+\sin \phi \gamma_5 \tau_2)\Psi$, $(iii) \Psi^\dagger (\cos \phi \tau_2 -\sin \phi \gamma_5 \tau_1)\Psi$
anticommute with the Hamiltonian, and provides a Dirac mass. Therefore, these three types of
Kondo singlet are energetically most competitive. These Kondo singlets will be the most pertinent ones,
when $v_+ < v_-$.

For case (ii), the Hamiltonian is
\begin{equation}\label{h1}
H_1=-i v_+ \gamma_0 \gamma_j \otimes \sigma_0 \otimes \tau_0 \partial_j +  g_- \gamma_0 \gamma_3
\mathbf{n} \cdot \boldsymbol \sigma \tau_3
\end{equation}Only the sublattice staggered Kondo singlet bilinears from Eq.~(\ref{kb2})
$(i) \Psi^\dagger \gamma_0 \gamma_3 ( \cos \alpha \tau_1 +\sin \alpha \tau_2 )\Psi$,
$(ii) \Psi^\dagger \gamma_0 \gamma_3 (\cos \phi \tau_1+\sin \phi \gamma_5 \tau_2 ) \Psi$,
$(iii) \Psi^\dagger \gamma_0 \gamma_3 (\cos \phi \tau_2 -\sin \phi \gamma_5 \tau_1) \Psi$
serve as the Dirac mass. These Kondo singlets will be the most pertinent ones, when $v_+ > v_-$.
The Kondo singlet bilinears from Eq.~(\ref{kb3}) never fully commute or anticommute
with the entire Hamiltonians even in these simple limits. Therefore, the Kondo singlets
from Eqs.~(\ref{kb1}) and (\ref{kb2}) are the energetically most competitive, and can lead to
paramagnetic Kondo insulator phases.

Since, in our microscopic model we have included an antiferromagnetic Kondo coupling,
rather than the frustrated Heisenberg couplings, it is natural to anticipate that the Kondo hybridization
will be preferred over the Peierls order. For strong enough magnetic frustration,
the Peierls order Eq.~(\ref{stsp}) will be possible. Thus the required critical value of the frustration
for nucleating Peierls order is enhanced by the Kondo coupling. However, we note that the above line
of reasoning for the mass generation is a weak coupling argument. This can always fail for a strong
coupling problem, where the fluctuation feedback is very important for energetics.

A well known example when the weak coupling argument fails is provided by the superfluid $^3He$,
where weak coupling argument always prefers fully gapped B phase. However, the gapless A
phase can become energetically competitive and even the bona-fide ground state,
after the spin fluctuation feedback effects are considered \cite{Anderson}. Therefore,
the charge and the current density waves and the Kondo singlets from Eq.~(\ref{kb3}) should not
be immediately discarded. In the following section we demonstrate how the AF, the Peierls and
the Kondo singlet bilinears can be rotated into each other through the chiral transformations.
This chiral relation will play an important role for the subsequent Berry phase considerations.

\section{Chiral rotation among the AF and competing singlet orders}
\label{chiral rotation}
For the simplicity we discuss the chiral rotation of competing orders for one and two species
of fermions separately. First we discuss the chiral relation between the Peierls and the AF orders
of one species. Subsequently we generalize this for the AF and the Kondo bilinears for two species.
\subsection{Rotation among the AF and the Peierls orders}
Since, the Peierls bilinears anticommute with the AF order parameter, we can combine these two orders
into the following O(5) chiral mass term\cite{TanakaHu,FisherSenthil,ChamonRyu}
\begin{equation}
\mathcal{O}_5=m \; \bar{\chi}\left[\cos \theta \exp (i\phi \gamma_5) + \sin \theta \mathbf{n} \cdot \boldsymbol \sigma \gamma_3\right]\psi.
\end{equation} Here, $\theta$ determines the relative strength between the singlet and the triplet orders,
and we can rotate two distinct orders into each other via unitary chiral transformations.
We can begin with the pure Peierls order of Eq.~(\ref{Ok}) and perform a spin dependent chiral transformation
\begin{equation}\label{ch1}
\chi \to e^{\frac{\theta}{2} \gamma_3 \mathbf{n} \cdot \boldsymbol \sigma} \: \: \bar{\chi}
\to e^{\frac{\theta}{2} \gamma_3 \mathbf{n} \cdot \boldsymbol \sigma},
\end{equation} which converts the $\mathcal{O}_K$ into $\mathcal{O}_5$. This is a unitary transformation,
as $\gamma_3$ is an anti-Hermitian matrix.

This chiral relationship among the Peierls and the AF bilinears is similar to the one dimensional problem,
where the single component Peierls and the AF order parameters are combined into a general O(4) chiral
mass term
\begin{equation}
\mathcal{O}_4=\bar{\chi}\left[\cos \theta i \gamma_5 + \sin \theta \mathbf{n} \cdot \boldsymbol
\sigma \right] \chi,
\end{equation}
and a spin dependent chiral transformation
\begin{equation}\label{ch2}
\chi \to e^{i \frac{\alpha}{2} \gamma_5 \mathbf{n} \cdot \boldsymbol \sigma} \: \: \bar{\chi}
\to e^{i \frac{\alpha}{2} \gamma_5\mathbf{n} \cdot \boldsymbol \sigma}
\end{equation}
causes the rotation between the distinct order parameters. When the 2+1-dimensional Dirac fermions
are chirally coupled to an O(5) mass, gradient expansion of the fermion determinant gives rise to a topological
Wess-Zumino-Witten (WZW) term\cite{TanakaHu,FisherSenthil}. If the spin Peierls order is integrated out,
the WZW term gives rise to the Berry phase of the sigma model field\cite{TanakaHu}.

\subsection{Rotation among the AF and the Kondo bilinears}
Given that the difference between the chiral charges is the generator of the Kondo singlet bilinears,
it is natural to ask if there is a way to rotate the AF order into the Kondo hybridization. We begin with
the dominant AF coupling
$$ g_- \bar{\Psi} \gamma_3 \mathbf{n} \cdot \boldsymbol \sigma \tau_3 \Psi$$ and perform the
perform the following chiral rotation
\begin{equation}
\Psi \to \exp \left( -\pi/4 \; \gamma_3 \mathbf{n} \cdot \boldsymbol \sigma \tau_0 \right)  \Psi ; \:
\bar{\Psi} \to \bar{\Psi} \exp \left( -\pi/4 \; \gamma_3 \mathbf{n} \cdot \boldsymbol \sigma \tau_0 \right)
\end{equation}
which converts the AF couplings to a Peierls bilinear
\begin{equation}\label{peierlsd}
g_- \bar{\Psi} \tau_3 \Psi.
\end{equation}
This is not surprising, as the AF order parameter can be chirally rotated to the Peierls order parameter
and vice versa. Now a subsequent species dependent, but spin independent transformation
\begin{eqnarray}\label{gamma0ch}
\Psi \to \exp \left( i \frac{\pi}{4} \gamma_0 \left \{ \cos \alpha \tau_1+ \sin \alpha \tau_2 \right \} \right ) \Psi,
\nonumber \\
 \bar{\Psi} \to \bar{\Psi} \exp \left ( -i \frac{\pi}{4} \gamma_0 \left \{ \cos \alpha \tau_1
 + \sin \alpha \tau_2 \right \} \right ),
\end{eqnarray}
rotates the Peierls bilinear of Eq.~(\ref{peierlsd}) to the following combination of the conventional Kondo
singlet bilinears
\begin{equation}
g_- \bar{\Psi} \gamma_0 \left(\sin \alpha \tau_1 -\cos \tau_2 \right) \Psi=g_- \Psi^\dagger \left(\sin \alpha
\tau_1 -\cos \tau_2 \right) \Psi
\end{equation}
appearing in Eq.~(\ref{kb1}). The $\gamma_5$ dependent Kondo hybridizations of Eq.~(\ref{kb1})
can be obtained through a subsequent chiral transformation shown in Eq.~(\ref{chdt}). Therefore,
the net chiral transformation for going from the AF coupling to the Kondo singlets of Eq.~(\ref{kb1})
is given by
\begin{eqnarray}\label{chkb1}
\Psi & \to & e^{ -\frac{\pi}{4} \; \gamma_3 \mathbf{n} \cdot \boldsymbol \sigma \tau_0 } \;
e^{ i \frac{\pi}{4} \gamma_0 \left \{ \cos \alpha \tau_1+ \sin \alpha \tau_2 \right \} } \;
e^{i \frac{\phi}{2} \gamma_5 \tau_3} \; \Psi, \nonumber \\
\bar{\Psi} & \to & \bar{\Psi} \; e^{i \frac{\phi}{2} \gamma_5 \tau_3} \; e^{- i \frac{\pi}{4} \gamma_0
\left \{ \cos \alpha \tau_1+ \sin \alpha \tau_2 \right \} } \; e^{ -\frac{\pi}{4} \; \gamma_3 \mathbf{n}
\cdot \boldsymbol \sigma \tau_0 }
\end{eqnarray}

We can similarly rotate the AF into the other Kondo singlet bilinears. For obtaining the $\gamma_j$
dependent bilinears we can replace $\gamma_0$ matrix in Eq.~(\ref{gamma0ch}) by $\gamma_j \gamma_5$,
which leads to
\begin{equation}
g_-\bar{\Psi} \gamma_j \gamma_5 \left(\sin \alpha \tau_1-\cos \alpha \tau_2 \right).
\end{equation}
These are the $\gamma_5$ dependent bilinears of Eqs.~ (\ref{kb2}) and (\ref{kb3}). Through a subsequent
chiral transformation as in Eq.~(\ref{chdt}), we can obtain the $\gamma_j$ dependent but $\gamma_5$
independent bilinears.
Therefore, the following transformations
\begin{eqnarray}\label{chkb2}
\Psi & \to & e^{ -\frac{\pi}{4} \; \gamma_3 \mathbf{n} \cdot \boldsymbol \sigma \tau_0 } \; e^{ i \frac{\pi}{4}
\gamma_j \gamma_5 \left \{ \cos \alpha \tau_1+ \sin \alpha \tau_2 \right \} } \; e^{i \frac{\phi}{2} \gamma_5
\tau_3} \; \Psi, \nonumber \\
\bar{\Psi} & \to & \bar{\Psi} \; e^{i \frac{\phi}{2} \gamma_5 \tau_3} \; e^{- i \frac{\pi}{4} \gamma_j \gamma_5
\left \{ \cos \alpha \tau_1+ \sin \alpha \tau_2 \right \} } \; e^{ -\frac{\pi}{4} \; \gamma_3 \mathbf{n}
\cdot \boldsymbol \sigma \tau_0 }, \nonumber \\
\end{eqnarray}
convert the AF term into the Kondo singlets of Eqs.~ (\ref{kb2}) and (\ref{kb3}). The chiral relationship among
the AF and the Kondo bilinears show that these distinct orders are part of a general chiral vacuum, with distinct
vacuum angles. For this reason they can appear as dual orders. The chiral relation will play a crucial role
in the discussion of the Berry phase in the magnetically disordered state.

\section{Berry phase and O(5) WZW term}
\label{Berry_phase_emergent}
Further insight into the role of the topological defects and the nature of the competing order is provided
by the Berry phase for the sigma model field. For the one dimensional problem we have shown in
Ref.~\onlinecite{Goswami1}, that a Berry phase term $-i \pi W[\mathbf{n}]$ is generated from the
fermion determinant, when the fermions scatter from the instanton configurations of the sigma model field.
This emergent Berry phase cancels the preexisting Berry phase of the spin half chain.
The cancelation of the Berry phase makes the sigma model gapped, and this is consistent
with the spin gap phase obtained via a bosonization analysis. The emergence of the Berry phase
from the fermion determinant can be shown through a chiral rotation method, where the AF coupling
is converted into the Peierls term by the following transformation
\begin{equation}
\psi \to e^{i \frac{\pi}{4} \gamma_5 \mathbf{n} \cdot \boldsymbol \sigma} \: \: \bar{\psi} \to e^{i \frac{\pi}{4}
\gamma_5\mathbf{n} \cdot \boldsymbol \sigma}
\end{equation}
The Berry phase term appears as a consequence of the chiral anomaly in odd spatial dimensions.
In one dimension the density of the Peierls order parameter exactly equals the instanton density of the
sigma model. Therefore, the cancelation of the Berry phase implies that the sum of the Peierls bilinears
from the two subsystem vanishes. But, the difference between the Peierls bilinears from two species
equals twice the instanton density. Therefore, akin to the two dimensions, a species staggered
Peierls order $\bar{\Psi} i\gamma_5 \tau_3 \Psi$ emerges as the competing order of the Kondo singlets.
Inside the Kondo assisted spin gapped phase, we do not have an expectation value for the
$\bar{\Psi} i\gamma_5 \tau_3 \Psi$, and only in the presence of a substantial amount of magnetic frustration,
the nucleation of the Peierls order will be possible.

We can also ask if the Berry's phase cancels inside the two dimensional disordered phase.
Given that the Berry's phase is proportional to the hedgehog invariant (see Eq.~(\ref{berry})),
which consists of the $\mathbf{n}$ field three times (see Eq.~(\ref{hedgehog})), we do anticipate such
a cancelation between the geometric phases emerging from the determinant of the two types of fermions.
We have mentioned before that the Berry phase is responsible for fixing the chiral angle $\phi$ of the
Peierls order into a $C_{3v}$ breaking pattern. Therefore, the cancelation of the Berry phase is consistent
with the species staggering of the chiral angle of the Peierls order as in Eq.~(\ref{stsp}). This argument
can be substantiated within the continuum theory, by evaluating the WZW term for the O(5) vector
formed out of the AF and the Peierls orders. We will also show that the WZW term can appear in the
presence of the $\gamma_5$ dependent Kondo singlets.

For simplicity we will set $g_+=0$, and first consider the following O(5) vector of the staggered
AF and the Peierls orders
\begin{eqnarray}
&&\bar{\Psi}\mathcal{V}_{5,1}\Psi=M \sum_{j=1}^{5} \; \bar{\Psi}\Gamma_j \mathcal{V}_{j,1}
\Psi \nonumber \\
&=&M\bar{\Psi} \bigg[\cos \theta \gamma_3 \otimes \tau_3 \mathbf{n} \cdot \boldsymbol \sigma
+ \sin \theta \cos \phi \mathbb{I} \otimes \tau_0 \nonumber \\&& + i \sin \theta \sin \phi \gamma_5
\otimes \tau_3 \bigg]\Psi.
\end{eqnarray}
We have denoted the five matrices that multiply the five components of the unit vector by $\Gamma_j$,
and $M$ is an overall scale for the amplitude or mass. The WZW term arises from the homotopy
classification $\Pi_{4}(S^4)=\mathbb{Z}$, and its evaluation is thoroughly described in
Ref.~\onlinecite{Abanov,TanakaHu,YaoLee}. The WZW term emerges when
\begin{eqnarray}\label{trace}
\mathrm{Tr}\bigg[\Gamma_a \gamma_\mu \Gamma_b \gamma_\nu \Gamma_c \gamma_\rho
\Gamma_d  \Gamma_e \bigg] \neq 0
\end{eqnarray}
Recall that the Greek indices are the space time indices 0,1,2 and $\gamma_5=i\gamma_0 \gamma_1
\gamma_2 \gamma_3$. The above trace can be nonzero if the product $\Gamma_a \Gamma_b
\Gamma_c \Gamma_d \Gamma_e$ is proportional to $\gamma_3 \gamma_5$. Since the AF bilinear
has the $\gamma_3$ matrix, the product of the two competing order components must produce
$\gamma_5$ matrix. This is indeed satisfied by the two components of the Peierls order parameter.
In addition the product of the two components must produce $\tau_3$ to absorb the $\tau_3$ of the
AF components. The staggered Peierls order parameter again satisfies this requirement.
The trace evaluation produces a five dimensional Levi-Civita symbol and the resultant WZW term
becomes
\begin{eqnarray}\label{WZW1}
S_{WZW,1}=2 \times \frac{-2\pi i \epsilon_{abcde} }{\mathcal{A}_4 } && \int^1_0 du \int d^2 x d\tau \;
\mathcal{V}_{a,1} \partial_u \mathcal{V}_{b,1}  \nonumber \\
&&\times \partial_\tau \mathcal{V}_{c,1} \partial_x \mathcal{V}_{d,1} \partial_y \mathcal{V}_{e,1}
\end{eqnarray}
where $\mathcal{A}_4=\frac{8\pi^2}{3}$ is the area of the hypersphere $S^4$. This is a level 2 WZW term,
which should be contrasted with the level 1 WZW term for the single species of fermion.
The topological relation between these distinct order also implies that the vortex core of the staggered
Peierls order carries the staggered AF order. In addition, we anticipate a non-Landau transition between
the staggered AF and the staggered Peierls phases.

If we have chosen the Peierls order without staggering we can define the O(5) vector
\begin{eqnarray}
&&\bar{\Psi}\mathcal{V}_{5,2}\Psi=M \sum_{j=1}^{5} \; \bar{\Psi}\Gamma_j \mathcal{V}_{j,2} \Psi \nonumber \\
&=&M\bar{\Psi} \bigg[\cos \theta \gamma_3 \otimes \tau_3 \mathbf{n} \cdot \boldsymbol \sigma
+ \sin \theta \cos \phi \mathbb{I} \otimes \tau_0 \nonumber \\&&
+ i \sin \theta \sin \phi \gamma_5 \otimes \sigma_0 \bigg]\Psi,
\end{eqnarray}
the trace in Eq.~(\ref{trace}) vanishes. Therefore, we do not have any WZW term for this quintuplet.
Since, the Berry phase for the nonlinear sigma model field is obtained by integrating out the Peierls
components from the WZW term, the net Berry phase vanishes.

Let us form the following quintuplet by combining the conventional Kondo singlets and the staggered AF
\begin{eqnarray}
&&\bar{\Psi}\mathcal{V}_{5,3}\Psi=M \sum_{j=1}^{5} \; \bar{\Psi}\Gamma_j \mathcal{V}_{j,2} \Psi \nonumber \\
&=&M\bar{\Psi} \bigg[\cos \theta \gamma_3 \otimes \tau_3 \mathbf{n} \cdot \boldsymbol \sigma + \sin \theta \cos \phi \gamma_0 \otimes \tau_1  \otimes \sigma_0 \nonumber \\&& + i \sin \theta \sin \phi \gamma_0 \otimes \tau_2
\otimes \sigma_0 \bigg]\Psi.
\end{eqnarray} We also set $v_+=0$. Due to the absence of the $\gamma_5$ matrix in the Kondo
hybridization terms, the trace in Eq.~(\ref{trace}) vanishes. Consequently, there is no WZW term for this quintuplet.
Therefore, a transition between the staggered AF and the conventional Kondo singlet phases without
the staggering, is conventional one. Generically this will be an O(3) transition, as anticipated in
Ref.~\onlinecite{Saremi}. A level 2 WZW term for a quintuplet with the conventional singlet can
be found, if a $\gamma_5$ matrix multiplies with the vector order parameter
\begin{eqnarray}
&&\bar{\Psi}\mathcal{V}_{5,4}\Psi=M \sum_{j=1}^{5} \; \bar{\Psi}\Gamma_j \mathcal{V}_{j,2} \Psi \nonumber \\
&=&M\bar{\Psi} \bigg[\cos \theta \gamma_3 \gamma_5 \otimes \tau_3 \mathbf{n} \cdot \boldsymbol \sigma
+ \sin \theta \cos \phi \gamma_0 \otimes \tau_1  \otimes \sigma_0 \nonumber \\&&
+ i \sin \theta \sin \phi \gamma_0 \otimes \tau_2 \otimes \sigma_0 \bigg]\Psi.
\end{eqnarray} In this case, instead of a staggered AF order, the O(3) vector field describes a staggered
quantum spin Hall order (in which the spin-Hall conductivities of the two species have a difference that is
quantized, but add up to zero). The competition between the Kondo singlet formation and the quantum
spin Hall state of the Kane-Mele model with $\mathbf{n}=(0,0,1)$ has recently been considered in
Ref.~\onlinecite{Feng}. However, the full vector order parameter of the spin Hall state has not been
considered yet. For the spin Hall ordered state, the role of the Peierls order in the $P_S$ phase,
is played by the s-wave superconducting state \cite{ChamonRyu,Grover,Herbut1,Moon,Chakravarty1}.
Thus $P_S$ to $P_L$ transition now occurs between the s-wave superconductor and the conventional
Kondo singlet states.

The level 2 WZW term for the transition between the staggered AF and the Kondo singlet phases
can be found, if we form the quintuplets with the help of the following four $\gamma_5$ dependent
Kondo hybridizations of Eqs.~(\ref{kb1}) and (\ref{kb2}),
$$ \bar{\Psi} \gamma_0 \left(\cos \phi \tau_{1/2} \pm \sin \phi \gamma_5 \tau_{2/1} \right)\Psi , \: \:
\mathrm{when} \: \: v_+=0,$$
$$ \bar{\Psi} \gamma_0 \gamma_3 \left(\cos \phi \tau_{1/2} \pm \sin \phi \gamma_5 \tau_{2/1} \right)\Psi,
\: \: \mathrm{when} \: \:  v_-=0.$$ Therefore, the vortices of these Kondo singlet states carry staggered AF order.
In addition, for these Kondo singlets, we anticipate a non-Landau transition
between the $AF_S$ and the $P_L$ phases.

\section{Conclusion and Outlook}
\label{conclusion}
We have presented an analysis of the half-filled Kondo-Heisenberg model on a honeycomb lattice, starting
from the antiferromagneticaly ordered insulating phase. We have described the AF order parameter
in terms of a quantum non-linear sigma model,
and emphasized the role of the topological defects and the associated
Berry's phase inside the magnetically disordered state. We have computed the induced Goldstone-Wilczek
currents for the fermion's chiral charge in the presence of the topological defects. The induced chiral current
for one species of spinful, four component Dirac fermion equals the skyrmion current of the AF background
(see Eq.~(\ref{sk1})), which in turn shows that the skyrmion defects carry the chiral charge (see Eq.~(\ref{comp1})).
The chiral charge has been shown to be the conjugate variable of many intervalley, translation symmetry
breaking competing orders such as the spin Peierls, the charge density wave and the current density wave states.

In the presence of the Kondo coupling there are two species of four component, spinful Dirac fermions.
Beginning with an order-parameter description of the AF insulating phase, where the staggered magnetizations
of the conduction fermions and the local moments are anti-aligned, we have found that the sum of the chiral
currents for the two species vanishes (see Eq.~(\ref{sum})). In contrast, the difference between the chiral
currents remains finite and equals twice the skyrmion current (see Eq.~(\ref{diff})). For this reason,
the skyrmion core carries the difference between the chiral charge. This result implies that the difference
between the chiral charges of the two species serves as the conjugate variable of various competing orders
(see Eq.~(\ref{compkondo1})).

Following this line of reasoning,  we have identified for the first time
different types of Kondo bilinears (see Eqs.~(\ref{kb1}), (\ref{kb2}), (\ref{kb3})) as the competing
orders of the antiferromagnetism and the spin Peierls orders. We have also clarified the lattice versions
of the possible Kondo singlet bilinears. Some of the competing Kondo singlets break discrete symmetries
of the honeycomb lattice, and these Kondo hybridizations will be bona-fide competing orders even
in the sense of local order parameter description of Landau-Ginzburg theory. Based on the anticommutation
relation of the participating matrices, we have pointed out a class of hybridizations that can generate bigger
gap at the Dirac points, and emerge as the dominant competing orders within a weak coupling argument.

Finally, we have shown how the Peierls and the Kondo hybridizations can be obtained from the AF order
through generalized chiral rotations (see Eqs.~(\ref{chkb1}), (\ref{chkb2})). This clearly demonstrates that the
AF and many of the competing singlet orders are parts of a general chiral vacuum, and are distinguished
by their chiral angles. Based on the chiral relationship, we have combined some of the anticommuting
competing orders into O(5) chiral masses for the Dirac fermions. For example, (i) the species staggered
AF and Peierls orders (ii) the species staggered AF  and some of the Kondo hybridizations of Eq.~(\ref{kb1})
and Eq.~(\ref{kb2}), can be combined as the O(5) chiral masses. For each of these combinations we have
shown the presence of a topological level 2 WZW term (see Eq.~(\ref{WZW1})). This topological term shows
that the participating orders of a quintuplet are indeed dual to each other and the core of topological defects
of one order carries the other competing order. Therefore, the vortices of the staggered Peierls and some
of the Kondo hybridizations possess the species staggered AF order. If the singlet orders are integrated
out, the WZW term can be reduced to a Berry phase term for the AF order, which involves the hedgehog defects.
The emergent Berry phase is twice the Berry phase of the local moments. The presence of the WZW
term suggests that depending on the nature of the Kondo singlets, the phase transition between the $AF_S$
and $P_L$ phases can fall outside the realm of Landau theory.

For calculational simplicity, we have chosen to work with the Dirac fermions. However, our procedure of the
induced current calculation can be extended for generic dispersions of the underlying fermions.
The calculation directly carries over for the quadratic band touching or any other Fermi points
in two dimensions. In Ref.~\onlinecite{FuSachdev}, a similar induced current calculation has been
performed for the tight binding fermions on a square lattice. For a generic dispersion and particularly
for the incommensurate metallic case, we do not expect the exact quantized relation between
the skyrmion current and the fermion's chiral current. If we consider the conduction fermions
of our model at away from the half-filling, the triangle diagram produces a chemical potential
dependent $j_{\psi,ch,\mu}$. Now both the sum and the difference of the chiral currents are finite,
and proportional to the skyrmion current. We just do not have the quantized factors of two or
zero anymore. However, a non vanishing $j_{d,ch,\mu}$ suggests that the skyrmion core still
carries the fluctuating Peierls and the Kondo hybridizations as competing orders. In three spatial
dimensions the role of the skyrmions as the static defects will be replaced by the hedgehogs. However,
we have to consider the chiral anomaly as discussed in Ref.~\onlinecite{Goswami1} for one dimension.

\section{acknowledgements}
This work has been supported in part through the National High Magnetic Field Laboratory by NSF Cooperative
Agreement No.\ DMR-0654118, the State of Florida, and the U. S. Department of Energy (P.G.),
and by the NSF Grants No.\ DMR-1006985, the Robert A.\ Welch Foundation Grant No.\ C-1411,
and the Alexander von Humboldt Foundation (Q.S.).  Q.S.\  also acknowledges the hospitality
of the the Karlsruhe Institute of Technology, the Aspen Center for Physics (NSF Grant No.\ 1066293),
and the Institute of Physics of Chinese Academy of Sciences.


\begin{thebibliography}{}

\bibitem{SiSteglich} Q.~Si and F.~Steglich,
Science \textbf{329}, 1161 (2010).

\bibitem{Lohneysen_rmp}
H.~v.~L\"{o}hneysen, A.~Rosch, M.~Vojta, and P.~W\"{o}lfle,
Rev.~Mod.~Phys. \textbf{79}, 1015 (2007).

\bibitem{Si_PhysicaB2006}Q. Si, Physica B \textbf{378}, 23 (2006);
Q. Si, Phys. Status Solidi B{\bf 247}, 476 (2010).

\bibitem{YamamotoSi_JLTP2010}S. J. Yamamoto, and Q. Si, J. Low Temp. Phys. \textbf{161}, 233 (2010).

\bibitem{Coleman_JLTP2010}P. Coleman and A. H. Nevidomskyy, J. Low Temp. Phys. \textbf{161}, 182 (2010).

\bibitem{Friedemann09}
S.~Friedemann {\it et al.}, Nat.~Phys. \textbf{5}, 465 (2009).

\bibitem{Custers10}J. Custers {\it et al.}, Phys. Rev. Lett. \textbf{104}, 186402 (2010).

\bibitem{Tokiwa09}Y. Tokiwa {\it et al.}, J. Phys. Soc. Jpn. \textbf{78}, 123708 (2009).

\bibitem{Custers12}
J. Custers {\it et al.}, Nature Mater. \textbf{11}, 189 (2012).

\bibitem{Aronson}
M. S. Kim and M. C. Aronson, Phys.~Rev.~Lett. \textbf{110}, 017201 (2013).

\bibitem{Canfield}
E. D. Mun {\it et al.}, Phys.~Rev.~B \textbf{87}, 075120 (2013).

\bibitem{Lohneysen}
V.~Fritsch {\it et al.}, arXiv:1301.6062.

\bibitem{Khalyavin}
D. D. Khalyavin {\it et al.},
Phys.~Rev.~B{\bf 87}, 220406(R) (2013).

\bibitem{Si-Nature} Q.~Si, S.~Rabello, K.~Ingersent, and J.~Smith,
Nature \textbf{413}, 804 (2001).

\bibitem{Coleman-JPCM}
P.~Coleman, C.~P\'{e}pin, Q.~Si, and R.~Ramazashvili,
J.~Phys.~Cond.~Matt. \textbf{13}, R723 (2001).

\bibitem{Paschen}
S.~Paschen {\it et al.},
Nature \textbf{432}, 881 (2004).

\bibitem{Shishido2005}H. Shishido {\it et al.}, J. Phys. Soc. Jpn. \textbf{74}, 1103 (2005).

\bibitem{Hertz}J. A. Hertz, Phys. Rev. B \textbf{14}, 1165 (1976).

\bibitem{Millis}A. J. Millis, Phys. Rev. B \textbf{48}, 7183 (1993).

\bibitem{Moriya}T. Moriya, {\it Spin Fluctutations in Itinerant Electron Magnetism} (Springer, Berlin, 1985).

\bibitem{SiPaschen}
Q.~Si and S.~Paschen, Phys. Status Solid B {\bf 250}, 425 (2013);
arXiv:1303.4141.

\bibitem{Hewson} A.\ C.\ Hewson, {\it The Kondo Problem to Heavy Fermions}
(Cambridge Univ.\ Press, Cambridge, 1993).

\bibitem{Elitzur} S. Elitzur, Phys. Rev. D \textbf{12}, 3978 (1975).

\bibitem{FradkinShenker} E. Fradkin, and S. H. Shenker, Phys. Rev. D \textbf{19}, 3682 (1979).

\bibitem{Burdin} S. Burdin, D. R. Grempel, and A. Georges, Phys. Rev. B \textbf{66}, 045111 (2002).

\bibitem{Senthiletal}
T. Senthil, M. Vojta and S. Sachdev, Phys.\ Rev.\ B \textbf{{69}}, 035111 ({2004}).

\bibitem{Pauletal} I. Paul, C. P\'{e}pin, and M. R. Norman, Phys. Rev. Lett. \textbf{98}, 026402 (2007).

\bibitem{Yamamoto2007}S. J. Yamamoto, and Q. Si, Phys. Rev. Lett. \textbf{99}, 016401 (2007).

\bibitem{Jones} T. Tzen Ong and B. A. Jones, Phys. Rev. Lett. \textbf{103}, 066405 (2009).

\bibitem{Haldane}F. D. M. Haldane, Phys. Rev. Lett. \textbf{50}, 1153 (1983).

\bibitem{Duncan} F. D. M. Haldane, Phys. Rev. Lett. \textbf{61}, 1029 (1988).

\bibitem{Chakravarty} S. Chakravarty, B. I. Halperin, and D. R. Nelson, Phys. Rev. B \textbf{39}, 2344 (1989).

\bibitem{MurthySachdev} G. Murthy and S. Sachdev, Nucl. Phys. B \textbf{344}, 557 (1990).

\bibitem{Arafune} J. Arafune, P. G. O. Freund, and C. J. Goebel, J. Math. Phys. \textbf{16}, 433 (1975).

\bibitem{Read} N. Read and S. Sachdev, Physical Review B \textbf{42}, 4568 (1990).

\bibitem{Tsvelik1994}A. M. Tsvelik, Phys. Rev. Lett. \textbf{72}, 1048 (1994).

\bibitem{Goswami1} P. Goswami and Q. Si, Phys. Rev. Lett. \textbf{107}, 126404 (2011).

\bibitem{Zachar}O. Zachar, and A. M. Tsvelik, Phys. Rev. B \textbf{64}, 033103 (2001).

\bibitem{SiPivovarov}E. Pivovarov, and Q. Si, Phys. Rev. B \textbf{69}, 115104 (2004).

\bibitem{Rajaraman} R. Rajaraman, {\it Solitons and Instantons} (North Holland, 1987).

\bibitem{Senthiletal2} T. Senthil {\it et al.}, Science \textbf{303}, 1490 (2004).

\bibitem{AffleckHaldane}I. Affleck, and F. D. M. Haldane, Phys. Rev. B \textbf{36}, 5291 (1987).

\bibitem{Hermele} M. Hermele, T. Senthil, M. P. A. Fisher, Phys. Rev. B \textbf{72}, 104404 (2005).

\bibitem{FisherSenthil} T. Senthil and M. P. A. Fisher, Phys. Rev. B \textbf{74}, 064405 (2006).

\bibitem{TanakaHu} A. Tanaka and X. Hu, Phys. Rev. Lett. \textbf{95}, 036402 (2005).

\bibitem{Chamon1} C. Y. Hou, C. Chamon, C. Mudry, Phys. Rev. Lett. \textbf{98}, 186809 (2007).

\bibitem{Jackiw} R. Jackiw and S.-Y. Pi Phys. Rev. Lett. \textbf{98}, 266402 (2007).

\bibitem{Wilczek1} J. Goldstone, and F. Wilczek, Phys. Rev. Lett. \textbf{47}, 986 (1981).

\bibitem{Tanaka1}A. Tanaka, J. Phys. C \textbf{320}, 012020 (2011).

\bibitem{FuSachdev} L. Fu, S. Sachdev, C. Xu, Phys. Rev. B \textbf{83}, 165123 (2011).

\bibitem{Deser}S. Deser, R. Jackiw, and S. Templeton, Phys. Rev. Lett. \textbf{48}, 975 (1982).

\bibitem{Semenoff} A. J. Niemi and G. W. Semenoff, Phys. Rev. Lett. \textbf{51}, 277 (1982).

\bibitem{Redlich}A. N. Redlich, Phys. Rev. D \textbf{29}, 2366 (1984).

\bibitem{Fradkin1} E. Fradkin, E. Dagotto, and D. Boyanovsky, Phys. Rev. Lett. \textbf{57}, 2967 (1986).

\bibitem{Jaroszewicz} T. Jaroszewicz, Phys. Lett. B \textbf{146}, 337 (1984).

\bibitem{Wilczek2} Y. H. Chen, and F. Wilczek, Int. J. Mod. Phys. B \textbf{3}, 117 (1989).

\bibitem{Abanov} A. G. Abanov and P. B. Wiegmann, Nucl. Phys. B \textbf{570}, 685 (2000).

\bibitem{HaldaneHall} F. D. M. Haldane, Phys. Rev. Lett. \textbf{61}, 2015 (1988).

\bibitem{Anderson} P. W. Anderson, and W. F. Brinkman, Phys. Rev. Lett. \textbf{30}, 1108 (1973).

\bibitem{ChamonRyu} S. Ryu, C. Mudry, C.-Y. Hou, and C. Chamon, Phys. Rev. B \textbf{80}, 205319 (2009).

\bibitem{YaoLee} H. Yao and D. H. Lee, Phys. rev. B \textbf{82}, 245117 (2010).

\bibitem{Saremi} S. Saremi, P. A. Lee, and T. Senthil, Phys. Rev. B \textbf{83}, 125120 (2011).

\bibitem{Feng}X. Y. Feng, J. Dai, C.-H. Chung, and Q. Si,
Phys. Rev. Lett. \textbf{111}, 016402 (2013).

\bibitem{Grover} T. Grover, and T. Senthil,  Phys. Rev. Lett. \textbf{100}, 156804 (2008).

\bibitem{Chakravarty1} C. H. Hsu, and S. Chakravarty, Phys. Rev. B 87, 085114 (2013).

\bibitem{Herbut1}C. K. Lu, and I. F. Herbut, Phys. Rev. Lett. \textbf{108}, 266402 (2012).

\bibitem{Moon} E. G. Moon, Phys. Rev. B \textbf{85}, 245123 (2012).





\end{thebibliography}
\end{document}